\documentclass[sigconf]{acmart}
\usepackage{subfigure}
\usepackage{algorithm}
\usepackage{algorithmic}
\usepackage{amsmath}
\usepackage{microtype}
\usepackage{balance}

\newcommand{\cI}{\mathcal{I}}
\newcommand{\cU}{\mathcal{U}}
\newcommand{\cC}{\mathcal{C}}
\newcommand{\cL}{\mathcal{L}}
\newcommand{\cK}{\mathcal{K}}

\newcommand{\R}{\mathbb{R}}

\DeclareMathOperator{\sign}{sign}

\AtBeginDocument{%
  \providecommand\BibTeX{{%
    \normalfont B\kern-0.5em{\scshape i\kern-0.25em b}\kern-0.8em\TeX}}}

\setcopyright{acmcopyright}
\copyrightyear{2021}
\acmYear{2021}
\acmDOI{10.1145/1122445.1122456}

\acmConference[CIKM '21] {Proceedings of the 30th ACM International Conference on Information and Knowledge Management}{November 1--5, 2021}{Virtual Event, Australia.}
\acmBooktitle{Proceedings of the 30th ACM Int'l Conf. on Information and Knowledge Management (CIKM '21), November 1--5, 2021, Virtual Event, Australia}
\acmPrice{15.00}
\acmISBN{978-1-4503-8446-9/21/11}
\acmDOI{10.1145/3459637.3481954}

\begin{document}
\fancyhead{}

\title{Sequential Search with Off-Policy Reinforcement Learning}


\author{Dadong Miao}
\author{Yanan Wang}
\author{Guoyu Tang}
\affiliation{%
  \institution{JD.com}
  \streetaddress{JD Building, No. 18 Kechuang 11 Street, BDA}
  \city{Beijing}
  \postcode{101111}
  \country{People's Republic of China}
}

\author{Lin Liu}
\author{Sulong Xu}
\author{Bo Long}
\affiliation{%
  \institution{JD.com}
  \streetaddress{JD Building, No. 18 Kechuang 11 Street, BDA}
  \city{Beijing}
  \postcode{101111}
  \country{People's Republic of China}
}

\author{Yun Xiao}
\author{Lingfei Wu}
\author{Yunjiang Jiang}
\affiliation{%
  \institution{JD.com Silicon Valley R\&D Center}
  \streetaddress{675 E Middlefield Road}
  \city{Mountain View}
  \state{CA}
  \postcode{94043}
  \country{USA}
}
\renewcommand{\shortauthors}{Dadong Miao, et al.}
\begin{abstract}
Recent years have seen a significant amount of interests in Sequential Recommendation (SR), which aims to understand and model the sequential user behaviors and the interactions between users and items over time. Surprisingly, despite the huge success Sequential Recommendation has achieved, there is little study on Sequential Search (SS), a twin learning task that takes into account a user's current and past search queries, in addition to behavior on historical query sessions.
The SS learning task is even more important than the counterpart SR task for most of E-commence companies due to its much larger online serving demands as well as traffic volume.

To this end, we propose a highly scalable hybrid learning model that consists of an RNN learning framework leveraging all features in short-term user-item interactions, and an attention model utilizing selected item-only features from long-term interactions.
As a novel optimization step, we fit multiple short user sequences in a single RNN pass within a training batch, by solving a greedy knapsack problem on the fly. 
Moreover, we explore the use of off-policy reinforcement learning in multi-session personalized search ranking. Specifically, we design a pairwise Deep Deterministic Policy Gradient model that efficiently captures users' long term reward in terms of pairwise classification error.
Extensive ablation experiments demonstrate significant improvement each component brings to its state-of-the-art baseline, on a variety of offline and online metrics.     
\end{abstract}





\maketitle



\begin{teaserfigure}
  \includegraphics[width=\textwidth]{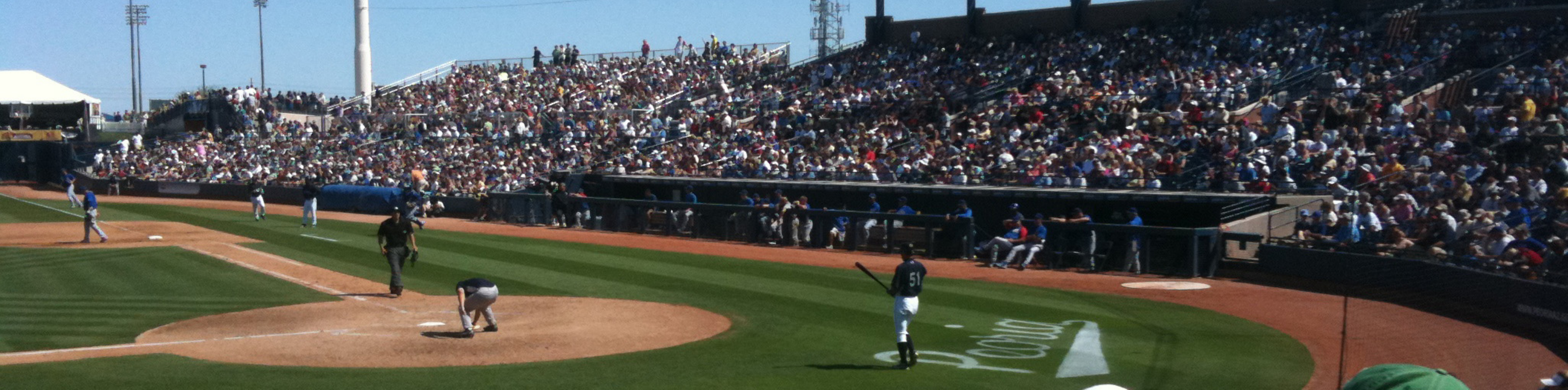}
  \caption{Seattle Mariners at Spring Training, 2010.}
  \Description{Enjoying the baseball game from the third-base
  seats. Ichiro Suzuki preparing to bat.}
  \label{fig:teaser}
\end{teaserfigure}

\section{Introduction}
\label{sec:introduction}

Over the past decade, neural network has seen an exponential growth in industrial applications. One of its most successful applications is in the area of search and recommendation. For instance, in the e-commerce domain, users often log onto the e-commerce platform with either vague or clear ideas of things they want to purchase. A recommender system proposes items to the user based on the items' popularity or the users' preference, inferred through past user interactions or other user profile information, while a search engine also takes into account the user provided search query as an input.

To capture a user's likely preference among the millions of items, recommendation systems have sought to leverage the user's historical interactions with the system. This is known as the Sequential Recommendation problem and has been studied thoroughly, especially after the neural network revolution in 2012, thanks to its remarkable modeling flexibility. The end result is that recommendation systems nowadays such as TikTok can deliver users' preferred content (measured in terms of post-recommendation interactions) with freakish accuracy, especially for frequent users. 

The analogous problem in the search domain, surprisingly, has been largely unexplored, at least in the published literature. Just like recommenders, search engines host billions of users. Many of these users may have interacted with the ranked results hundreds of times, through clicks in general, and also adding to cart or purchase actions in the e-commerce domain. Furthermore, unlike with recommendation systems, search engine users also leave a trail of their search queries, each of which narrows down the space of potential result candidates and can help the search engine better understand specific areas of the user's interests.

To help close this gap in the search domain, we propose a new class of learning tasks called Sequential Search. To the best of our knowledge, this is the first paper to formally introduce this new problem domain. Despite the simlarity with the well-known Sequential Recommendation problems, Sequential Search has its own sets of opportunities and challenges. First and foremost, the provision of a query in the current session significantly restricts the candidate result pool, making it feasible to approach from a ranking perspective, instead of retrieval only. At the same time, the quality of the ranked results also depends heavily on the quality of the retrieval phase. Similarly, The presence of historical queries issued by the user in principle provides much more targeted information about his or her preferences with respect to different search intents. On the flip side, however, changing queries break the continuity of the result stream presented to the user, leading to sparsity of latent ranking signals, as well as rendering personalization secondary to semantic relevance.

The present paper is an attempt to take on the aforementioned opportunities and challenges at the same time. Similar to DIEN \cite{zhou2019deep} in the Sequential Recommendation literature, we experiment with a combination of attention and RNN network to capture users' historical interactions with the system. Besides the modeling consideration stated in \cite{zhou2019deep}, such as capturing users' interest evolution, our hybrid sequential approach is also motivated by the desire to update the model incrementally during online serving, for which RNN is naturally more efficient than attention. In addition, due to the difficulty of gradient propagation over long sequences of recurrent network,  we specialize the attention component (Figure~\ref{fig:din}) to deal with long term interactions (up to 1 year) with limited number of categorical feature sequences, while apply the RNN network only to near term interactions (within the past 30 days) with the full set of features. 

Because of the multiple queries involved in the user interaction sequence, our training data takes on a special nested format (Figure~\ref{fig:user_session_tsv}) which can be viewed as a 4-dimensional array. By contrast, users in sequential recommendation problems can be treated as having a single session, albeit extended over a long period of time potentially. To further deal with the unevenness of user sequence lengths within a training minibatch, we devise a novel knapsack packing procedure to merge several short user sequences into one, thereby significantly reducing computational cost during training.

Finally to optimize for long term user experience and core business metrics, we build a deep reinforcement learning model naturally on top of the RNN framework. As is typically done, the users are treated as a partially observable environment while the recommender or search engine itself plays the role of the agent, of which the model has full control. Following ideas similar to the Deep Deterministic Policy Gradient network \cite{lillicrap2015continuous}, we introduce \textbf{S}equential \textbf{S}ession \textbf{S}earch \textbf{D}eep \textbf{D}eterministic \textbf{P}olicy \textbf{G}radient network, or S3DDPG, that optimizes a policy gradient loss and a temporal difference loss at the same time, over a continuous action space represented by the RNN output embedding as well as the agent prediction score (See Figure~\ref{fig:ddpg}). 

One difficulty in applying reinforcement learning in the search context is that the environment is essentially static. Online model exploration is expensive in terms of core business metrics, especially if the model parameters need to be explored and adjusted continuously. Fortunately our model can be trained completely offline, based only on the logged user interactions with the search results. In particular, we do not introduce any (offline) simulator component in the reinforcement learning training cycle, including the popular experience replay technique \cite{mnih2015human}. This significantly reduces model complexity compared to similar work \cite{zhao2017deep} in the Sequential Recommendation domain. Thanks to the presence of the temporal difference loss, S3DDPG also does not seem to require an adjustment of the underlying trajectory distribution, unlike the policy gradient loss only approach taken by \cite{chen2019top}.

In a word, we summarize our main contributions as follows:
\begin{itemize}
    \item We formally propose the highly practical problem class of Sequential Search, and contrast it with the well-studied Sequential Recommendation.
    \item We design a novel 4d user session data format, suitable for sequential learning with multiple query sessions, as well as a knapsack algorithm to reduce wasteful RNN computation due to padding.
    \item We present an efficient combination of RNN framework leveraging all features in near-term user interactions, and attention mechanism applied to selected categorical features in the long-term.
    \item We further propose a Sequential Session Search Deep Deterministic Policy Gradient (S3DDPG) model that outperforms other supervised baselines by a large margin.
    \item We demonstrate significant improvements, both offline and online, of RNN against DNN, as well as our proposed S3DDPG against RNN, in the sequential search problem setting.
\end{itemize}

All source code used in this paper will be released for the sake of reproducibility. 
\footnote{Source code will available at https://github.com/xxxx (released upon publication)}

\section{Related Work}
\label{sec:related_work}

\subsection{User Behavior Modeling}
\label{subsec:rw:user}

User behavior modeling is an important topic in industrial ads, search, and recommendation system. A notable pioneering work that leverages the power of neural network is provided by Youtube Recommendation \cite{covington2016deep}. User historical interactions with the system are embedded first, and sum-pooled into fixed width input for downstream multi-layer perception. 

Follow-up work starts exploring the sequential nature of these interactions. Among these, earlier work exploits sequence models such as RNN \cite{hidasi2015session}, while later work starting with \cite{li2017neural} mostly adopts attention between the target example and user historical behavior sequence, notably DIN \cite{zhou2018deep} and KFAtt\cite{liu2020kalman}. 

More recently, self-attention \cite{kang2018self} and graph neural net \cite{wu2019session,pang2021heterogeneous} have been successfully applied in the sequential recommendation domain.



\subsection{RNN in search and recommendation}
\label{subsec:rw:rnn}

While attention excels in training efficiency, RNN still plays a useful role in settings like incremental model training and updates. 
Compared to DNN, RNN is capable of taking the entire history of a user into account, effectively augmenting the input feature space. Furthermore, it harnesses the sequential nature of the input data efficiently, by constructing a training example at every event in the sequence, rather than only at the last event \cite{zhang2014sequential}. Since the introduction of Attention in \cite{vaswani2017attention}, however, RNN starts to lose its dominance in the sequential modeling field, mainly because of its high serving latency. 

We argue however RNN saves computation in online serving, since it propagates the user hidden state in a forward only manner, which is friendly to incremental update. In the case when user history can be as long as thousands of sessions, real time attention computation can be highly impractical, unless mitigated by some approximation strategies \cite{drachsler2008personal}. The latter introduces additional complexity and can easily lose accuracy. 

Most open-source implementations of reinforcement learning framework for search and recommendation system implicitly assume an underlying RNN backbone \cite{chen2019top}. The implementation however typically simplifies the design by only feeding a limited number of ID sequences into the RNN network \cite{zhao2019deep}. 

\cite{zhang2019deep} contains a good overview of existing RNN systems in Search / Recommendations. In particular, they are further divided into those with user identifier and those without. In the latter case, the largest unit of training example is a single session from which the user makes one or more related requests. While in the former category, a single user could come and go multiple times over a long period of time, thus providing much richer contexts to the ranker. It is the latter scenario that we focus on in this paper. To the best of our knowledge, such settings are virtually unexplored in the search ranking setting.


\subsection{Deep Reinforcement Learning}
\label{subsec:rw:rl}

While the original reinforcement learning idea was proposed more than 3 decades ago, there has been a strong resurgence of interest in the past few years, thanks in part to its successful application in playing Atari games \cite{mnih2013playing}, DeepMind's AlphaGo \cite{silver2017mastering} and in text generation domains \cite{chen2019reinforcement,gong2019reinforcement}. Both lines of work achieve either super-human level or current state-of-the-art performance on a wide range of indisputable metrics.  

Several important technical milestones include Double DQN \cite{van2016deep} to mitigate over-estimation of Q value, and \cite{schaul2015prioritized}, which introduces experience replay. However, most of the work focuses on settings like gaming and robotics. We did not adopt experience replay in our work because of its large memory requirement, given the billion example scale at which we operate. 

The application in personalized search and recommendation has been more recent. Majority of the work in this area focuses on sequential recommendation such as \cite{zhao2018deep} as well as ads placement within search and recommendation results \cite{zhao2021dear}.

An interesting large scale off-policy recommendation work is presented in \cite{chen2019top} for youtube recommendation. They make heuristic correction of the policy gradient calculation to bridge the gap between on-policy and off-policy trajectory distributions. We tried it in our problem with moderate offline success, though online performance was weaker, likely because our changing user queries make the gradient adjustment less accurate.

Several notable works in search ranking include \cite{hu2018reinforcement} which takes an on-policy approach and \cite{xu2020reinforcement} which uses pairwise training examples similar to ours. However both works consider only a single query session, which is similar to the sequential recommendation setting, since the query being fixed can be treated as part of the user profile. In contrast, our work considers the user interactions on a search platform over an extended period of time, which typically consist of hundreds of different query sessions.

\label{subsec:rw:rl_search}


\section{Method}
\label{sec:method}
We introduce the main model architecture in this section. Each subsection forms the foundation for the next one. Section~\ref{sec:din} describes an attention based network inspired by \cite{zhou2018deep} that exploits the correlation between the current ranking task and the entire historical sequence of items for which the user has expressed interest. Section~\ref{sec:rnn} details an elaborate Recurrent Neural Net backbone that can efficiently handle a batch of uneven sized sequences, and how it is used to capture more recent user interactions with the search engine. The output embedding of the attention network is simply fed as an input into every timestamp of the recurrence. Finally we discuss how to build an actor-critic style reinforcement learning model on top of the RNN structure in section~\ref{sec:ddpg}. 

\subsection{Attention for Long-Term Session Sequence}
\label{sec:din}
As mentioned in section~\ref{subsec:rw:user}, the attention network in the Deep Interest Network (DIN) model is a natural way to incorporate user history into personalized recommendation. To adapt to the search ranking setting, we introduce Search Ranking Deep Interest Network (DIN-S), which makes the following adjustments on top of DIN:
\begin{itemize}
    \item Query side features are introduced alongside the focus item features, to participate in attention with historical item sequence.
    \item To account for the possibility that the current query request is not correlated with any of the user's past actioins, a zero score is appended to the regular attention scores before getting the softmax weights. This is illustrated by the zero attention unit in Figure~\ref{fig:din}.
\end{itemize}https://www.overleaf.com/project/5fed74e5fac2600586cb62da
\begin{figure}
    \centering
    \includegraphics[width=\linewidth]{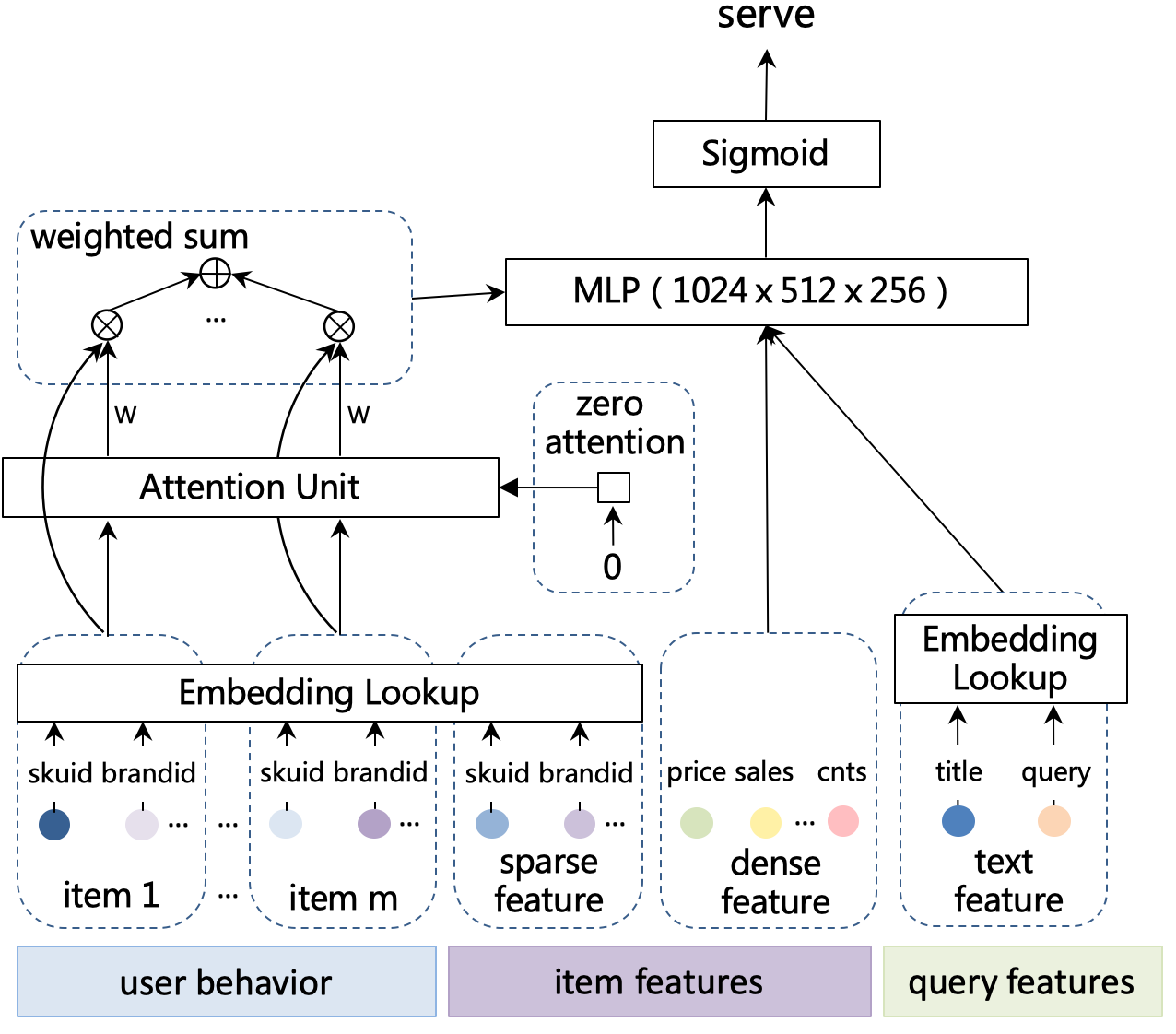}
    \centering
    \caption{DIN-S architecture.}
    \label{fig:din}
\vspace{-5pt}
\end{figure}

The overall architecture of DIN-S is outlined in Diagram~\ref{fig:din}. Due to the nature of algorithmic iterations within an industrial setting, DIN-S is not only one of our quality comparison baselines, but also one of the major components in our proposed final architecture.
We have also tried DIEN \cite{zhou2019deep} and other follow up works. Despite the better results reported in papers, we found little incremental improvement in our own systems. However our design of the RNN backbone model shares some similarity with the approach in DIEN, and indeed both ours and the DIEN work use attention and RNN together. A key difference, however, is that our RNN training data and algorithmic design uses all features available in previous interactions by the user, including both item or query/user one-sided as well as two-sided features. By contrast, the RNN (GRU) in DIEN appears to be just an extension of the co-existing Attention network, taking mostly item-side only categorical features.

\subsection{RNN for Near-Term Sequential Search}
\label{sec:rnn}
\begin{figure}
    \centering
    \includegraphics[width=\linewidth]{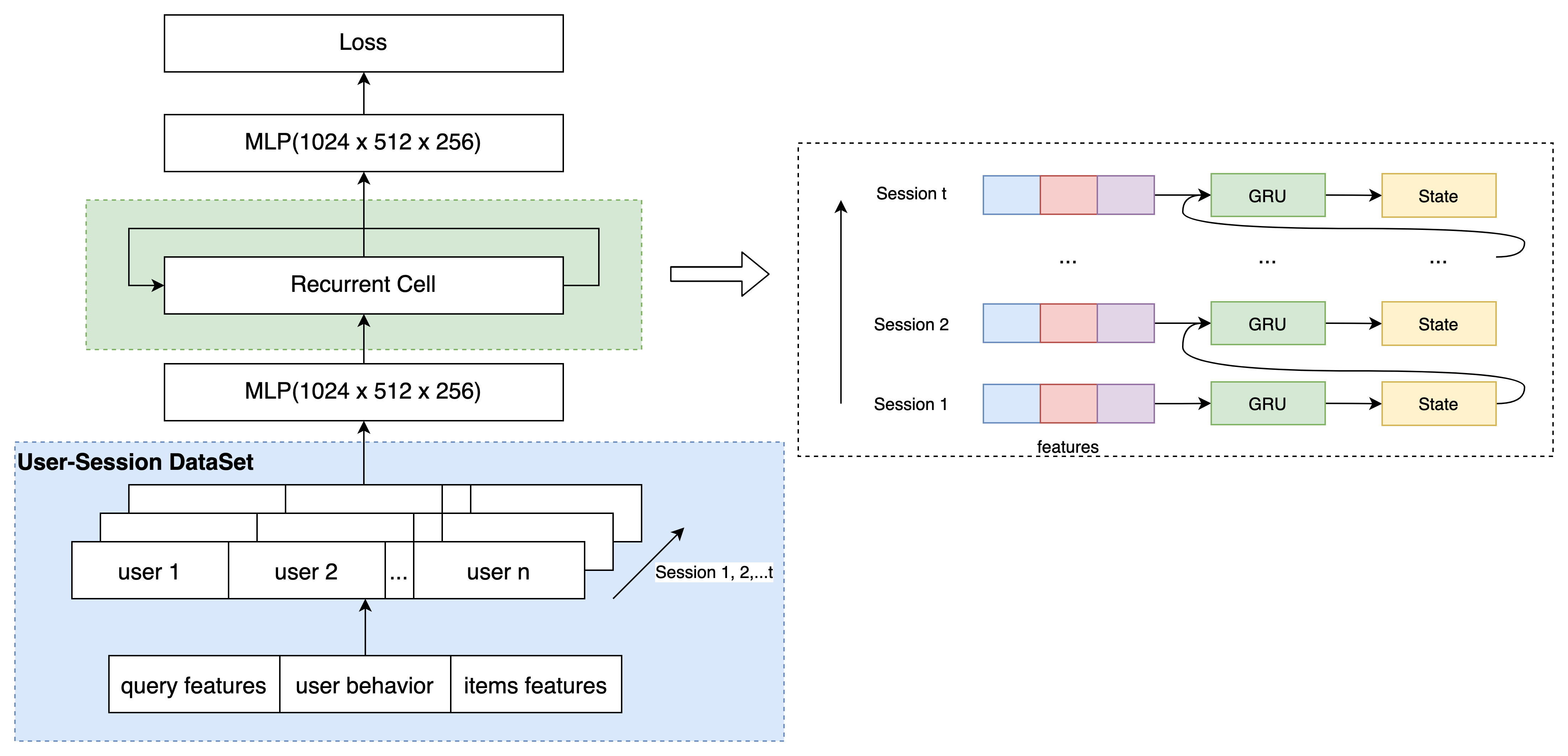}
    \centering
    \caption{RNN architecture.}
    \label{fig:rnn}
\end{figure}

In order to compare with the baseline DIN-S (attention + MLP) model fairly and conveniently, we build a so-called RNN backbone that can wrap around any base model architecture. The logic is outline in the bottom half of Diagram~\ref{fig:rnn}. To summarize, for any base model $M$, the RNN backbone introduces a new feature vector $H_t$, the hidden state, concatenated to the output of $M$. The output of each time iteration of the RNN model is another vector $H_{t+1}$, which serves both as the input to downstream networks, as well as the hidden state input for the next time iteration. 

\subsubsection{Contiguous Session-Based Data Format}
\label{subsubsec:data_format}
While DIN-S can be trained in a pointwise / pairwise fashion, our implementation of RNN tries to pool all relevant information together in the data by 
\begin{itemize}
    \item arranging all items within a query session in a single training example. In our case we used tsv (tab-separated values) format. Thus the number of columns in each row is variable, depending on the number of items under the session.
    \item placing all query sessions belonging to the same user contiguously to ensure they are loaded altogether in a mini-match.
\end{itemize}
\begin{figure}
    \centering
    \includegraphics[width=\linewidth]{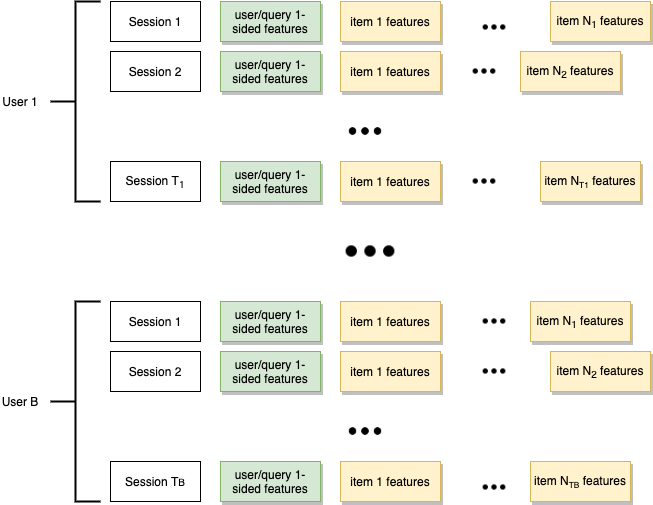}
    \centering
    \caption{User Session Input Format. $B$ stands for batch size. Each row represents a single TSV row in the input data. The numbers of columns = number of query features + num of items $\times$ number of item features.}
    \label{fig:user_session_tsv}
\vspace{-5pt}
\end{figure}
As illustrated in Figure~\ref{fig:user_session_tsv}, each mini-batch thus contains a 4d tensor $B$ whose elements are indexed by $(u, t, i, f)$, which stand for users, sessions (time-ordered), items, and features respectively. We assume that features are all dense or have been converted to fixed width dense format, through either embedding sum-pooling or 
attention-pooling from the DIN-S base model. We will use $B_{u, t}$ to denote the 2d slices of $B$ containing all $(i, f)$ values. 

\subsubsection{RNN Model Architecture}
Let $\omega_{u, t}, H_{u, t}$ stand for the regular output and hidden state output of the RNN network for user $u$ and session $t$. The RNN network can thus be described by a function $F$ with the following signature
\begin{align} \label{eq:rnn_kernel}
    (\omega_{u, t+1}, H_{u, t+1}) = F(B_{u, t + 1}, H_t).
\end{align}
This is the most general form of an RNN network. All RNN variants such as LSTM, GRU obey the above signature of $F$.

Recall now $B_{u, t}$ is a 2d tensor, with dimension given by (the number of items, number of features). The same is true of the output tensor $\omega_{u,t}$. The hidden state $H_{u, t}$ however has \textbf{no item dimension}: it is a fixed width vector for a given user after a given session. For simplicity, our choice of $F$ simply computes $H_{u, t}$ as a weighted average of the output $\omega_{u, t}$. More precisely,
\begin{align}
    \omega_{u, t+1}, S_{u, t+1} &= \rm{GRU}(\omega_{u, t}, S_{u, t}) \\
    H_{u, t+1, f} &= \frac{1}{|\cC_{u, t}|} \sum_{j \in \cC_{u, t}} \omega_{u, t+1, j, f}. 
\end{align}
Here $\cC_{u, t}$ stands for the set of items in user $u$'s session $t$ that were purchased. Those sessions without purchases are excluded from our training set, since under the above framework, 
\begin{enumerate}
    \item the user hidden state would not be updated;
    \item the final pairwise training label contains no information.
\end{enumerate} 
The vast majority of the remaining sessions contain exactly 1 purchased item.

This completes the description of the RNN evolution of the state vector under a listwise input format, where all items in a session are used. For training efficiency, however, we adopt a pairwise setup, where 2 items are sampled from each query session, and exactly one of them has been purchased. Thus we can think of each session as consisting of exactly 2 items. Since the hidden state is a weighted average over only the purchased items, pairwise sampling preserves all the information for the hidden state vector in a single RNN step, provided the session contains only a single purchased item, which is more than $90\%$ of the cases.

Lastly, the RNN model outputs a single logit $\eta_{u, t, i}$ for each item $i$ chosen within the user session $(u, t)$, by passing the RNN output vector $\omega_{u, t, i} \in \R^d$ of dimension $d$ through a multi-layer perception $P$ of dimensions $[1024, 256, 64, 1]$:
\begin{align} \label{eq:rnn_output}
    \eta_{u, t, i} = P(O(u, t, i)),  \quad P: \R^d \to \R.
\end{align}
The corresponding label is a binary indicator $\lambda_{u, t, i} \in \{1, 0\}$, which denotes whether the item was purchased or not.

\subsubsection{Pairwise Loss} \label{subsec:rnn:pairwise}
Unlike clicks or mouse hover actions, each page session in e-commerce search typically receives at most one \textbf{purchase}. Thus we are confronted with severe positive and negative label imbalance. To address this problem, we choose pairwise loss in our modeling, which samples a purchased item from the current session at random, and matches it with a random item that is viewed or clicked but not purchased. 

The exact sampling procedure is described in Algorithm~\ref{alg:pairwise_sampling}. Note that as long as the session is non-empty, the procedure will always output a pair. There are occasional edge cases when all items are purchased, in which case we output two purchased items. Alternatively, such perfect sessions can be filtered from the training set.

\begin{algorithm}[t]
\caption{Pairwise sampling from a query session.}
\label{alg:pairwise_sampling}
\begin{algorithmic}[1]
\REQUIRE{a list of $N > 0$ labels: $\lambda_1, \ldots, \lambda_N \in \{0, 1\}$}
\ENSURE{two indices $1 \leq a, b \leq N$, s.t. $\lambda_a = \max \lambda_i$ and $\lambda_b = \min \lambda_i$}
\STATE Compute $\lambda_{\min} := \min_i \lambda_i$ and $\lambda_{\max} := \max_i \lambda_i$.
\STATE Construct the list of admissible pairs $A := \{(a, b) \in [N]^2: \lambda_a = \lambda_{\max}, \lambda_b = \lambda_{\min}\}$ 
\STATE Output a uniformly random element $(a, b)$ from $A$.
\end{algorithmic}
\end{algorithm}
The final loss function on an input session $(u, t)$ is given by the following standard sigmoid cross entropy formula:

\begin{align} \label{eq:logloss}
    &\cL(B_{u, t}, \lambda_{u, t}) = -\lambda_{u, t} \log \sigma(\eta_{u, t}) - (1-\lambda_{u, t}) \log (1 - \sigma(\eta_{u, t}))
\end{align}
where 
\begin{itemize}
\item $B_{u, t}$ stands for all the features available to the model for a given user session $(u, t)$.
\item $\lambda_{u, t} := \lambda_{u, t, a, b} = \frac{\lambda_{u, t, a}}{\lambda_{u, t, a} + \lambda_{u, t, b}} \in \{0, 1\}$, depending on whether a purchase was made on item $a$ or $b$ within the user session $(u, t)$. 
\item $\eta_{u, t} := \eta_{u, t, a, b} = \eta_{u, t, a} - \eta_{u, t, b}$ is simply the difference between the model outputs for the two items $a$ and $b$, which can be interpreted as the log-odds that the purchase was made on the first item.
\item $a, b$ are a pair of random item indices within the current session, chosen according to Algorithm~\ref{alg:pairwise_sampling}, where item $a$ is purchased while item $b$ is not.
\item $\sigma(\eta_{u, t})$ transforms the pairwise logit $\eta_{u, t}$ through the sigmoid function $\sigma: x \mapsto (1 + e^{-x})^{-1}$, and can be interpreted as the model predicted probability that item $a$ is purchased, given exactly one of item $a$ and $b$ is purchased.
\end{itemize}

\subsubsection{Knapsack Sequence Packing}
Since the numbers of historical sessions vary widely across different users, the naive implementation of the above 4d representation can be computationally quite wasteful due to excessive zero padding. We thus adopt a knapsack strategy (Algorithm~\ref{alg:knapsack_rnn}) to fit multiple short user session sequences into the maximum length seen in the current mini-batch. 

\begin{algorithm}[t]
\caption{Parallel RNN via Knapsack Packing}
\label{alg:knapsack_rnn}
\begin{algorithmic}[1]
\REQUIRE{a list of $N$ (user, session) indices: $\cI = \{(u_1, 1), \ldots, (u_1, T_1), \ldots, (u_n, T_n)\}$}
\REQUIRE{input feature vectors associated with each (user, session) pair: $\{B_{u, t} \in \R^D: (u, t) \in \cI\}$}
\REQUIRE{an expensive RNN kernel $\tilde{O}: \R^{2D} \to \R^{2D}$}
\ENSURE{efficient computation of $\{\omega_{u, t} := O(B_{u, t}): (u, t) \in \cI\}$}
\STATE Apply the greedy knapsack strategy (Algorithm~\ref{alg:greedy_knapsack}) to get a mapping $m: (u, t) \mapsto (u', t')$, as well as the 2d array $S := \{S_{u', t'}\}$ that encodes the starting positions of the subsequences.
\STATE Construct a new input features $B'$ according to $B'_{u', t'} = B_{u, t}$.
\STATE Zero pad the missing entries of $B'$, for vectorized processing.
\STATE Compute $\omega' := O'(B', S)$ for all packed users in parallel.
\STATE Rerrange $\omega'_{u', t'}$ into the original user sequences $\omega_{u, t}$ via the inverse map $m^{-1}: (u', t') \mapsto (u, t)$.
\end{algorithmic}
\end{algorithm}
To break down Algorithm~\ref{alg:knapsack_rnn}, we introduce a few terminologies:
\begin{definition}
For a given RNN kernel $\tilde{O}: \R^D \times \R^D \to \R^D \times \R^D$, its associated \textbf{sequence map} $O: \R^{D \times T} \to \R^{D \times T}$, $(B_1, \ldots, B_T) \mapsto (\omega_1, \ldots, \omega_T)$ is given inductively by 
\begin{align*}
    (\omega_1, H_{u, 1}) &:= \tilde{O}(B_{u, 0}, H_{u, 0}) \\
    (\omega_{t+1}, H_{u, t+1})  &:= \tilde{O}(\omega_t, H_{u, t}) \quad \text{for } t \leq T -1.
\end{align*}
The initial hidden state is typically chosen to be the all zero vector: $H_{u, 0} = \vec{0}$.
\end{definition}

Note that after applying the knapsack packing Algorithm~\ref{alg:greedy_knapsack}, the maximum length of all the sequences stays the same. The total number of sequences, however, is reduced, by an average factor of 20x. As a result, some new sequence now contains multiple old sequences, arranged contiguously from the left. In such cases, we do not want the hidden states to propagate across sequences. Thus we introduce the following extended RNN sequence map that takes into account the old sequence boundary information:
\begin{definition}
Given an RNN kernel $\tilde{O}$ as above, and a 2d indicator array $\{S_t \in \{0, 1\}: 1 \leq t \leq T\}$ denoting the starting positions of sub-sequences within each user sequence, the \textbf{boundary-aware sequence map} 
\begin{align*}
    O': \R^{D \times T} \times  \{0, 1\}^{D \times T} \to \R^{D \times T}, \quad (B_1, \ldots, B_T) \mapsto (\omega_1, \ldots, \omega_T)
\end{align*} 
is defined via the following inductive formula
\begin{align*}
    (\omega_1, H_{u, 1}) &:= \tilde{O}(B_{u, 0}, H_{u, 0}) \\
    (\omega_{t + 1}, H_{u, t + 1}) &:= 
    \begin{cases}
      \tilde{O}(B_{u, t}, H_{u, 0}) & \text{if  $S_{t+1} = 1$ }\\
      \tilde{O}(\omega_t, H_{u, t}) & \text{otherwise}
    \end{cases}  
\end{align*}
\end{definition}
 

Overall the session knapsack strategy saves about 20x compute and speed up CPU training time by about 3x. Note that during online serving, knapsack is not needed since we deal with one user at a time.
\begin{algorithm}[t]
\caption{Greedy Knapsack Sequence Packing.}
\label{alg:greedy_knapsack}
\begin{algorithmic}[1]
\REQUIRE{A nonempty index set $\cI := \{(u_1, 1), \ldots, (u_1, T_1), \ldots, (u_n, T_n)\}$.}
\ENSURE{an index map $m: \cI \to \cU' \times [T']$, where $|\cU'| \leq n$ is the packed user index set and $T' \leq \max_i T_i$.}
\ENSURE{a 2d array $S_{u', t'}$ indicating the start positions of subsequences in each packed user sequence.}  
\STATE Set $T' := \max_{i=1}^n T_i$.
\STATE Initialize $U := [n] \setminus \{i\}$.
\STATE Initialize the list of knapsacks $\cK \leftarrow []$.
\STATE Initialize $S_{u', t'}$ to the all $0$ 2d array.
\WHILE {$U \neq \emptyset$}
    \STATE Pop a longest user sequence from $U$, say $u_j$.
    \IF {$T_j + \sum_{k \in \cK_i} T_k < T'$ for some $i \leq |\cK|$}
        \STATE Define $m(j, \ell) := (i, \ell + \sum_{k \in \cK_i} T_k)$ for $\ell < T_j$.
        \STATE Set $S_{i, \sum_{k \in \cK_i} T_k} \leftarrow 1$.
        \STATE Append $j$ to the end of $K_i$.
    \ELSE
        \STATE Append $[j]$ to the end of $\cK$.
        \STATE Define $m(j, \ell) := (|\cK|, \ell)$
        \STATE Set $S_{|\cK|, 1} \leftarrow 1$.
    \ENDIF
\ENDWHILE

\end{algorithmic}
\end{algorithm}
\subsection{DDPG for Near-Term Future Sessions}
\label{sec:ddpg}

While attention and RNN are capable of leveraging past sequential data, they fall short of predicting or optimizing future user behavior several steps in advance. This is not surprising because the former are essentially trained in a supervised approach, where the target is simply the next session. To optimize trajectories of several future sessions, we naturally turn to the vast repertoire of reinforcement learning (RL) techniques. 

As mentioned in Section~\ref{subsec:rw:rnn}, unlike the vast majority of RL literature in search and recommendation, our trajectory of agent (ranker) / environment (user) interaction is not confined within a single query session. Instead the user continues to type new queries, over a span of weeks or months. Thus the environment changes from one session to the next. However a key assumption here is that the different manifestations of the environment (user) share an underlying preference theme, as a single user's shopping tastes are strongly correlated across multiple shopping categories or intents.

Another important difference between our sequential session setup and the single session setup in other works is that each step of S3DDPG needs to rank a list of tens or hundreds of items, rather than just picking the top K from the remaining candidate pool. Due to the combinatorial explosion associated with ranking tasks, it becomes infeasible to treat the set of permutations of the items as our action space. Instead we take the vector output of the RNN network, along with the actor network prediction, as the action, which lives in a continuous space.


\subsubsection{S3DDPG Network}

\begin{figure}
    \centering
    \includegraphics[scale=0.5]{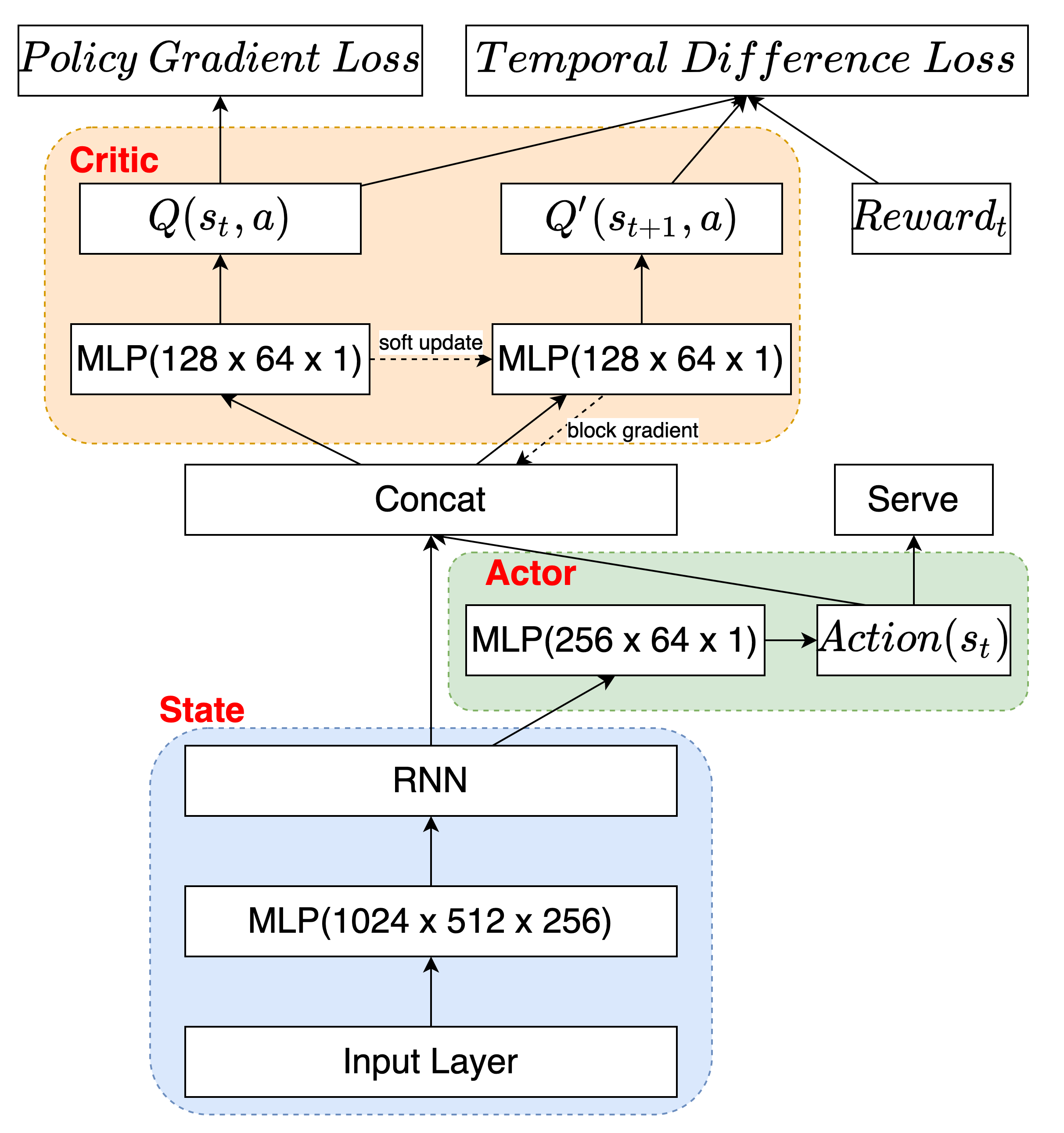}
    \centering
    \caption{S3DDPG architecture.}
    \label{fig:ddpg}
\end{figure}

Finally we come to our reinforcement-learning based ranking framework, which is depicted in Diagram~\ref{fig:ddpg}. The bottom half of the network consists of the RNN structure described in the previous subsection. The reinforcement learning part takes the regular RNN output (i.e., non-hidden state related) as the input, and is similar to the actor/critic framework. We closely follow the logic of DDPG network \cite{lillicrap2015continuous}.

The actor network has the same structure as the final MLP layers in RNN that takes the intermediate embedding to per-item logit. The latter is thus used also as the final ranking score for each item within a single query session. 

The critic network (also known as the Q-network) is a separate multi-layer perceptron, $Q: \R^d \to \R$, taking \textbf{a pair of RNN outputs} $\omega_{u, t, a}, \omega_{u, t, b} \in \R^d$ to a single scalar logit. 
\begin{align*}
    q_{u, t} := Q(\omega_{u, t, a}, \omega_{u, t, b}) \in \R.
\end{align*}

$Q$ is introduced here to approximate the following maximal cumulative discounted long term reward:
\begin{align*}
    q_{u, t} \sim \sup_{\eta_{u, t}, \ldots, \eta_{u, T}} \sum_{s = t}^T \gamma^{s - t} r(\eta_{u,  t}, \lambda_{u, t}).
\end{align*}

Here the supremum is taken over all trajectories starting at session $t$, and the reward $r(\eta_{u,t},\lambda_{u, t}) =: r_{u, t}$ is simply given by the opposite of the sigmoid cross entropy loss $\cL(B_{u, t}, \lambda)$ (See \eqref{eq:logloss}):
\begin{align} \label{eq:reward_definition}
    r(\eta_{u, t}, \lambda_{u, t}) = \lambda_{u, t} \log \sigma(\eta_{u, t}) + (1 - \lambda_{u, t}) \log \sigma(1 - \eta_{u, t}).
\end{align}
The time horizon $T$ itself is also random in general.

To summarize, we have introduced three networks and their associated output layers so far
\begin{itemize}
    \item $\omega_{u, t, a}, \omega_{u, t, b} \in \R^d$ are the output vectors of the RNN network for the chosen item pair.
    \item $\eta_{u, t} = P(\omega_{u, t, a}) - P(\omega_{u, t, b})$ is the scalar output of the actor network, which has the interpretation of log-odds of the first item being purchased.
    \item $q_{u, t} = Q(\omega_{u, t, a}, \omega_{u, t, b})$ is the scalar output of the critic (Q) network for the pair.
\end{itemize}

The critic (Q) network differs significantly from the actor network $P$ in that the input consists of pairs of items. Thus unlike $\eta_{u, t}$, it is not anti-symmetric under swapping of the item pair.

It is interesting to note that the original supervised loss function $\cL(\eta, \lambda)$ has been re-purposed as the reward in the Q-network. The actual loss functions are defined next.

\subsubsection{Loss Functions}
There are two loss functions in the S3DDPG framework. The first of these two, the temporal difference (TD) loss, is well-known since the first DQN paper \cite{mnih2013playing}. It aims to enforce the Bellman's equation on the Q-values:
\begin{align} \label{eq:bellman}
    q_{u, t} = \sup_{\eta_{u, t}} r(\eta_{u, t}, \lambda_{u, t}) + \gamma q_{u, t + 1}.
\end{align}
Here $\gamma$ is a discount factor, which is set to $0.8$ throughout our experiments. The associated TD loss would then be
\begin{align} \label{eq:dqn_td_loss}
    \cL^{\text{DQN}}_{\text{TD}} (B_{u, t}, \lambda_{u, t}) := \sum_{u \in \cU} \sum_{t=1}^{T - 1} (q_{u, t} - \sup_\eta \left\{r_t(\eta, \lambda) - \gamma q_{u, t+1}\right\})^2.
\end{align}
Here $\cU$ stands for all the users in the training data, and $T$ implicitly depends on the choice of $u$. 

As mentioned in Section~\ref{sec:ddpg}, however, our action space is either combinatorially explosive ($10!$), or continuous $\R^d$. Thus it is unclear how to compute the supremum on the right hand side. Instead we simply drop the supremum operator and consider the following weakened Bellman equation
\begin{align} \label{eq:weak_bellman}
    q_{u, t} = r(\eta_{u, t}, \lambda_{u, t}) + \gamma q_{u, t + 1}, \quad q_{u, T} = 0.
\end{align}
The TD loss thus aims to minimize the sum-of-square error between the two sides of the equation above:
\begin{align} \label{eq:td_loss}
    \cL_{\text{TD}}(B_{u, t}, \lambda_{u, t}) := \sum_{u \in \cU} \sum_{t=1}^{T - 1} (q_{u, t} - r_{u, t} - \gamma q_{u, t + 1})^2.
\end{align}
The problem with the above weakened TD loss \eqref{eq:td_loss} is that by itself, it is under-specified. Indeed, $r_{u, t} = r(\eta_{u, t}, \lambda_{u, t})$ can take on any (negative) value without affecting $\cL_{\text{TD}}$, since the extra degrees of freedom in $q_{u, t}$ can easily compensate for its wild moves. By contrast, the original TD Loss (for DQN) \eqref{eq:dqn_td_loss} eliminates this extra degree of freedom by taking the supremum over all actions $\eta_{u, t}$. 

To make the training loss fully specified, we thus introduce a second loss term, the policy gradient (PG) loss, which seeks to maximize the cumulative Q-value over the RNN and critic network model parameters. 
\begin{align}
    \cL_{\text{PG}}(B_{u, t}, \lambda_{u, t}) := \sum_{u \in \cU} \sum_{t=1}^T q_{u, t}, \quad q_{u, t} = Q(O(B_{u, t})).
\end{align}
where recall $q_{u, t} = Q(O(B_{u, t, a}), O(B_{u, t, b}))$ for the chosen positive / negative item pair. Note that since the actor network also depends on the RNN network parameters, the PG loss also indirectly optimizes over the action space. Furthermore, since $q_{u, t}$ are very closely tied with the supervised reward function $r_t$, by maximizing $q_{u, t}$, we are implicitly also maximizing the original supervised reward.



As is standard in DQN and DDPG, we also add the so-called target Q-network \cite{mnih2015human}, denoted by $\tilde{Q}$, that differs from the original Q-network only by one time-step, which is useful for stabilizing its learning. In other words, the exact weight updates are given by,
\begin{align}
    Q &\leftarrow Q + \alpha \nabla_Q (\sum_{t = 1}^{T - 1} Q(\omega_t) - \gamma \tilde{Q}(\omega_{t+1}) - r_t) \\
    \tilde{Q} &\leftarrow Q ,
\end{align}
where $\alpha$ is the effective learning rate that depends on the actual 1st order optimizer used.



We have also tried two versions of the actor networks, but the difference in evaluation metrics is small (about 0.04\% in session AUC), thus was discarded for simplicity and training efficiency.

Another important way S3DDPG differs from traditional DDPG implmementation is the relation between the two losses and weight updates. In the original proposal \cite{lillicrap2015continuous}, the actor and critic network weights are updated separately by the PG and TD losses:
\begin{align*}
    P \leftarrow P + \alpha \nabla_P \cL_{\text{PG}}, \qquad Q \leftarrow Q + \alpha \nabla_Q \cL_{\text{TD}}.
\end{align*}
However we cannot get the model to converge under this gradient update schedule. Instead we simply take the sum of the two losses $\cL_{\text{PG}} + \cL_{\text{TD}}$, and update all the network weights according to
\begin{align*}
    O, P, Q \leftarrow \alpha \nabla_{O, P, Q} (\cL_{\text{PG}} + \cL_{\text{TD}}).
\end{align*}

\section{Online Incremental Update}
\begin{figure}
    \centering
    \includegraphics[width=\linewidth]{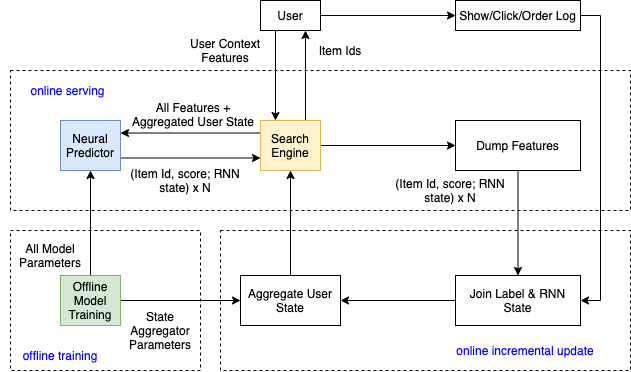}
    \caption{Real-Time Incremental Update Pipeline.}
    \label{fig:incremental_update}
\vspace{10pt}
\end{figure}

To capitalize on the underlying RNN modeling framework, we perform incremental update when the model is served online, so that the most recent user interactions can be captured by the model to update the user states. The overall architecture and its relation to offline training is summarized in Figure~\ref{fig:incremental_update}. The offline trained model can be divided into two sets of network parameters: 
\begin{itemize}
    \item The user state aggregation network takes the hidden states associated with all the items in the session, along with their corresponding labels, and perform average pooling to obtain a fixed size updated user state. If the session contains no purchase action, we do not update the user state.
    \item The remaining network take in the usual input features, along with the user state, to output predictions for each item.
\end{itemize}
The first of these is sent to an online incremental update component. While the latter goes directly to the neural network scorer.

The online serving component is roughly divided into three modules. At the center is the search engine itself, which is in charge of distributing and receiving features. 

When a user types in a query, the associated user context features, including query text, user's basic profile information, as well as user's historical actions, are all sent to the search engine. The search engine then relays this information to the neural network predictor, which in turn computes the predicted scores as well as the hidden state for each item, all of which are sent back to the search engine. Finally if the user makes any purchase in the current session, the new user state is updated to be the average of hidden states from the purchase items. 

\section{Experiment}
\label{sec:experiment}
\subsection{Evaluation Setup}


\subsubsection{Training Data Generation}
We collect 30 days of training data from our in-house search log. Table~\ref{tab:in-house-data} summarizes its basic statistics. The total number of examples in DIN-S (pre-RNN) training is 200m, while under the RNN/S3DDPG data format, we have 6m variable length sessions instead. While the majority of users only have a single session, the number of sessions per user can go as high as 100. This makes our knapsack session packing algorithm~\ref{alg:knapsack_rnn} a key step towards efficient training. 
\begin{table}[htbp]
\centering
\caption{In-house data statistics.}
\small
\begin{tabular}{c|c|c|c}
\hline
statistics & mean & minimum & maximum \\
\hline
Number of unique users & 3788232 & - & - \\
\hline
sessions per user &	13.42 & 1 & 113 \\
\hline
items per session & 26.97 & 1 & 499 \\
\hline
Features per (query, item) & 110 & - & - \\
\hline
\end{tabular}
\label{tab:in-house-data}
\end{table}

A characteristic of e-commerce search sessions is the huge variance in browsing depth (number of items in a session). In fact, some power user can browse content up to thousands of items within a single session. 
The short sessions (such as the minimum number of 2 items in the table) are due to lack of relevant results.

Each DIN-S training example consists of a single query and a single item under the query session. To leverage users' historic sequence information, the data also includes the item id, category id, shop id, and brand id of the historical sequence of clicked / purchased / carted items by the current user. The sequence is truncated at a maximum length of 500 for online serving efficiency. 

For RNN and S3DDPG, each example consists of a pair of items under the same query. In order to keep the training data compact, i.e., without expanding all possible item pairs, the training data adopts the User Session Input format (Section~\ref{subsubsec:data_format}). To ensure all sessions under a user are contained within each minibatch, and ordered chronologically, the session data is further sorted by session id as primary key and session timestamp as secondary key during the data generation mapreduce job.


During training, a random pair of items is sampled from each session, with one positive label (purchased) and one negative label (viewed/clicked only). Thus each minibatch consists of $\sum_{u=1}^B |S_u|$ item pairs, where $S_u$ stands for the set of all sessions under user $u$ and $B$ is the minibatch size, in terms of number of users. 

\subsubsection{Offline Evaluation}
We evaluate all models on one day of search log data beyond the training period. For RRNN and S3DDPG, however, 
we also include $N-1$ days prior to the last day, for a total of $N = 30$ days. The first $29$ days are there to build the user state vector only. Their labels are needed for user state aggregation during RNN forward evolution. Only labels from the last day sessions are used in the evaluation metrics, to prevent any leakage between training and validation.

\subsubsection{Offline Evaluation Metrics}
While cross entropy loss \eqref{eq:reward_definition} and square loss \eqref{eq:dqn_td_loss} are used during training of S3DDPG, for hold-out evaluation, we aim to assess the ability of the model to generalize forward in time. Furthermore even though the training is performed on sampled item pairs, in actual online serving, the objective is to optimize ranking for an entire session worth of items, whose number of can reach the hundreds. Thus we mainly look at session-wise metrics such as Session AUC or NDCG. Session AUC in particular is used to decide early stopping of model training: 

\begin{align} \label{eq:session_auc}
    \text{Session AUC}(\eta, \lambda) := \sum_{u = 1}^B \sum_{t = 1}^{|S_u|} \rm{AUC}(\eta_{u, t}, \lambda_{u, t}),
\end{align}
where $\eta_{u, t}$ denotes the list of model predictions for \textbf{all items} within the session $(u, t)$ and $\lambda_{u, t}$ the corresponding binary item purchase labels. This is in contrast with training, where $\eta_{u, t}$, $\lambda_{u, t}$ denote predictions and labels for a randomly chosen positive / negative item pair. 

The following standard definition of ROC AUC is used in \eqref{eq:session_auc} above. For two vectors $\boldsymbol{p}, \boldsymbol{t} \in \R^n$, where $t_i \in \{0, 1\}$:
\begin{align*}
\rm{AUC}(\boldsymbol{p}, \boldsymbol{t}) := \frac{\sum_{1 \leq i < j \leq n} \sign(p_i - p_j) \sign(t_i - t_j)}{n(n-1) / 2},
\end{align*}
where $\sign(x) = x / |x|$ for $x \neq 0$ and $\sign(0) = 0$.


NDCG is another popular metric in search ranking, intended to judge full page result relevance \cite{distinguishability2013theoretical}. It again takes the model predictions (which can be converted into ranking positions) as well as corresponding labels for all items, and compute a position-weighted average of the label, normalized by its maximal possible value:
\begin{align*}
    \rm{NDCG}(\boldsymbol{p}, \boldsymbol{t}) = \sum_{i=1}^n \frac{2^{t_i} - 1}{\log_2(i + 1)} / \sum_{i=1}^{\sum_j t_j} \frac{2^{t_i} - 1}{\log_2(i + 1)}.
\end{align*}


\subsubsection{Online Metrics}
For e-commerce search, there are essentially three types of core online metrics. 
\begin{itemize}
    \item GMV stands for gross merchandise value, which measures the total revenue generated by a platform. Due to the variation of A/B bucket sizes, it is often more instructive to consider GMV per user.
    \item CVR stands for conversion rate and essentially measures the number of purchases per click. Again this is averaged over the number of users.
    \item CTR is simply click-through rate, which measures number of clicks per query request. We do not consider this metric in our online experiments since it is not directly optimized by our models.
\end{itemize}

\subsection{Evaluation Results}

\begin{table}[htbp] 
\centering
\caption{Offline Metrics}
\begin{tabular}{c|c|c}
\hline
Model name & Session AUC & NDCG \\
\hline
DNN & 0.6765 & 0.5104 \\
DIN-S & 0.6875 & 0.5200 \\
RNN & 0.6915 & 0.5272 \\
S3DDPG & 0.6968 & 0.5307 \\
\hline
\end{tabular}
\label{tab:offline_metrics}
\end{table}
We present both Session AUC and NDCG for the 4 models listed in Table~\ref{tab:offline_metrics}. The DNN baseline simply aggregates the user sequential features through sum-pooling, all of which are id embeddings. The successive improvements are consistent between the two session-wise metrics: RNN improved upon DIN-S by about $0.4\%$ in Session AUC and $0.7\%$ in NDCG, while S3DDPG further improves upon RNN by another $0.5\%$ in Session AUC and $0.4\%$ in NDCG. The overall gain of S3DDPG is around a full $1\%$ in either metrics from the DIN-S baseline, and $2\%$ from the DNN baseline.

Table~\ref{tab:user_group_evaluation} highlights the gain of S3DDPG over the myopic RNN baseline on a variety of user subsets. For instance, along the dimension of users' past session counts, S3DDPG shows a significantly stronger performance for more seasoned users in both validation metrics. Another interesting dimension is whether the current query belongs to a completely new category of shopping intent compared to the users' past search experience. Users who issue such queries in the evaluation set are labeled ``Category New Users". Along that dimension, we see that S3DDPG clearly benefits more than RNN from similar queries searched in the past.

\begin{figure}
    \centering
    \includegraphics[width=\linewidth]{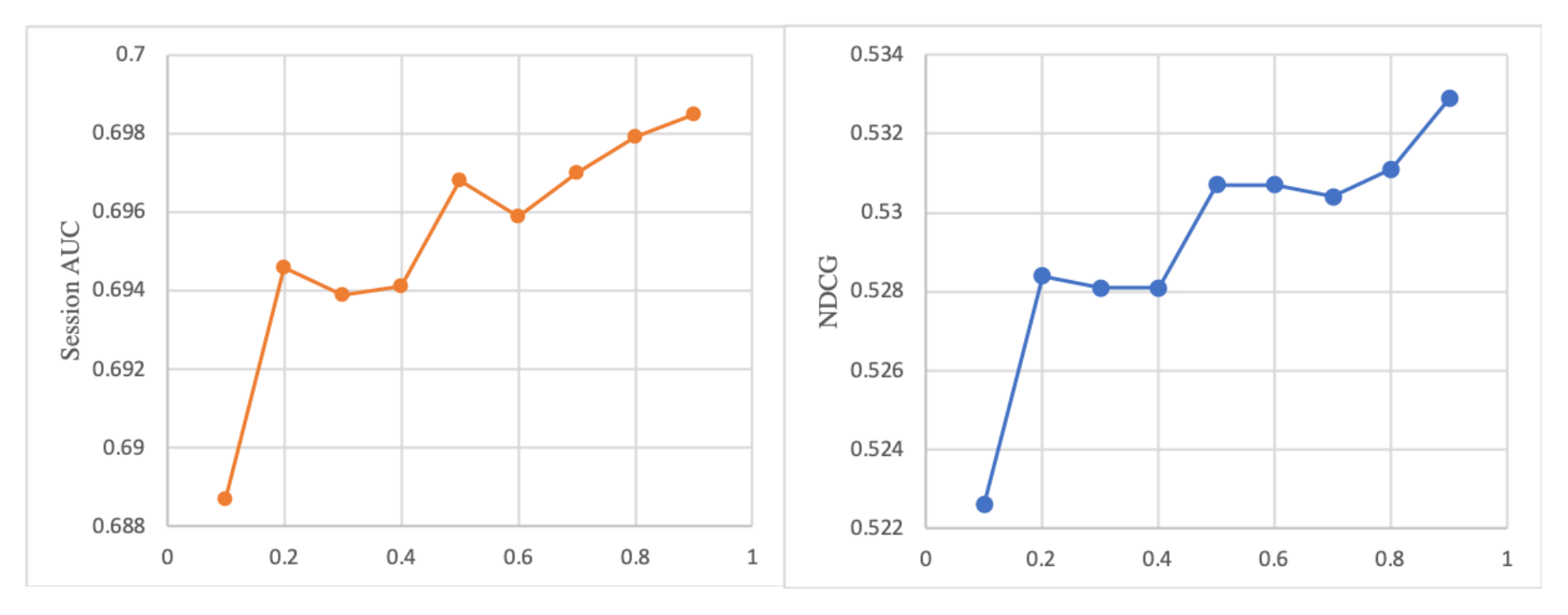}
    \caption{influence of hyperparameter $\mu$}
    \label{fig:hyperparameter}
\end{figure}

While there are a number of hyperparameters associated with reinforcement learning models in general, the most important one is arguably the discount factor $\gamma$ parameter. We choose $\gamma = 0.8$ for all our S3DDPG experiments, since we found little improvement when switching to other $\gamma$ value. Another important parameter specific to actor-critic style architecture is the relative weight $\mu$ between PG loss and TD loss. Interestingly, as we shift weight from TD to PG loss (increasing $\mu$), there is a noticeable trend of improvement in both AUC and NDCG, as shown in Figure~\ref{fig:hyperparameter}. This suggests the effectiveness of maximizing the long term cumulative reward directly, even at the expense of less strict enforcement of the Bellman equation through the TD loss. When $\mu$ is set to $1$, however, training degenerates, as the Q-value optimized by PG loss is not bound to the actual reward (cross entropy loss) any more.

\begin{figure}
    \centering
    \includegraphics[width=\linewidth]{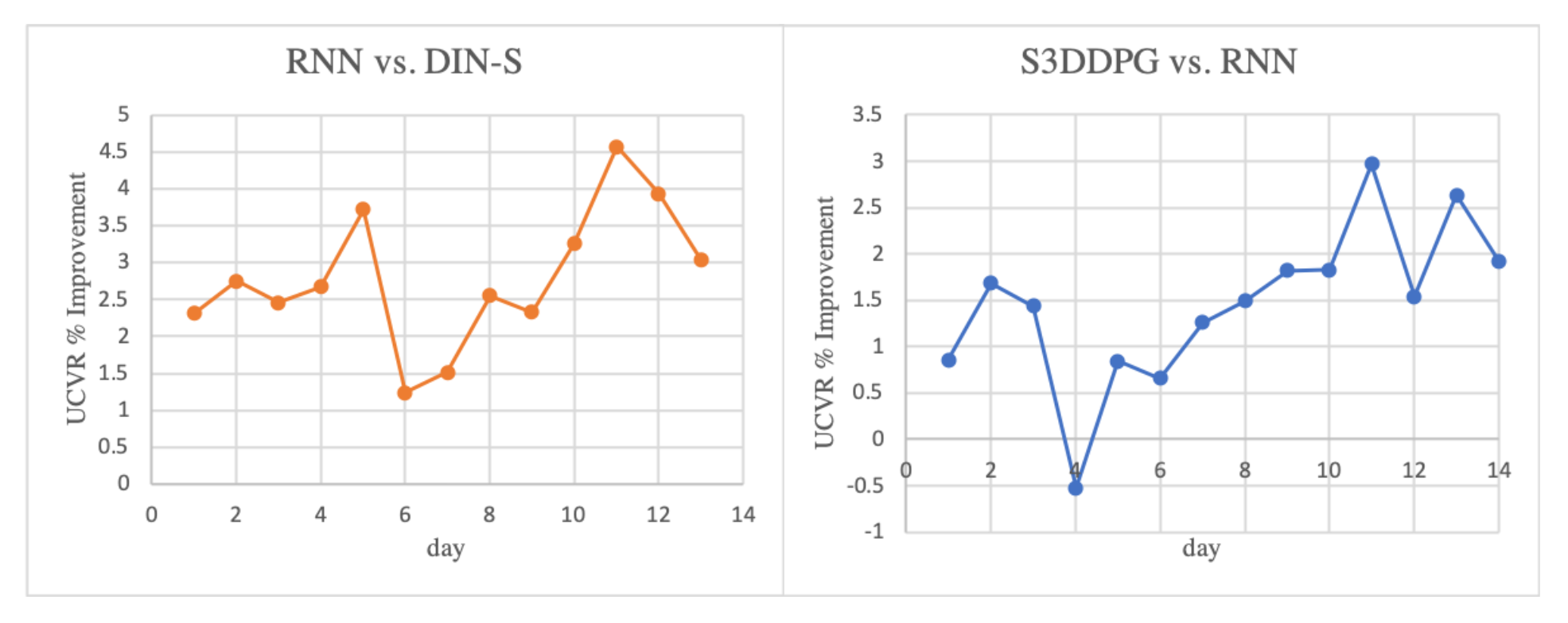}
    \caption{Daily UCVR \% improvement for online A/B tests over 14 days.}
    \label{fig:daily_ucvr}
\end{figure}

Finally we conduct 3 sets of online A/B tests, each over a timespan of 2 weeks. The overall metric improvements are reported in Table~\ref{tab:online_metrics}. The massive gain from DNN to DIN-S is expected, since the DNN baseline, with sum-pooling of the sequential features, is highly ineffective at using the rich source of sequential data. Nonetheless we also see modest to large improvements between RNN and DIN-S and between S3DDPG and RNN respectively, in all core business metrics. Figures~\ref{fig:daily_ucvr} present the daily UCVR metric comparison for the last 2 sets of A/B tests.  Aside from a single day of traffic variation, both RNN and S3DDPG show consistent improvement over their respective baselines. 

\begin{table}[htbp]
\centering
\caption{S3DDPG vs RNN for different user evaluation groups}
\begin{tabular}{c|c|c}
\hline
User Group  & Session AUC & NDCG \\
\hline
Past Session Count < 5  & +0.6534\% & +0.2418\% \\
Past Session Count >= 5  & +0.9544\% & +0.9092\%\\
Category New Users & +0.3195\% & +0.1412\% \\
Category Old Users & +0.6584\% & +0.6688\% \\
\hline
\end{tabular}
 \label{tab:user_group_evaluation}
\end{table}

\begin{table}[htbp]
\centering
\caption{Online Metrics}
\begin{tabular}{c|c|c|c}
\hline
Model pairs & GMV/user & CVR/user & RPM \\
\hline
DIN-S vs DNN & +4.05\%(1e-3) & +3.51\%(1e-3) & +4.08\%(5e-3) \\
RNN vs DIN-S & +0.60\%(5e-3) & +1.58\%(9e-3) & +0.49\%(8e-3) \\
S3DDPG vs RNN & +1.91\%(8e-3) & +0.78\%(1e-2) & +1.94\%(9e-3) \\
\hline
\end{tabular}
 \label{tab:online_metrics}
\end{table}
\vspace{-10pt}

\label{sec:industrial}


\label{sec:case}

\section{Conclusion}
\label{sec:conclusion}

We propose a previously unexplored class of machine learning problems, Sequential Search, that closely parallels Sequential Recommendation but comes with its own set of challenges and modeling requirements. We present three successively more advanced sequential models, DIN-S, RNN, and S3DDPG, to exploit the rich source of historical interactions between the user and the search platform. Practical techniques such as user session training data format, knapsack rearrangement of session sequence, and online incremental serving are discussed extensively. Finally systematic offline experiments on a large scale industrial dataset are performed, showing significant incremental improvement between all successive model pairs. These results are further validated with substantial gains in online A/B tests, showing that principled modeling of users' sequential behavior information can significantly improve user experience and personalized search relevance.


\bibliographystyle{ACM-Reference-Format}
\balance
\bibliography{references}
\citestyle{acmauthoryear}

\end{document}